\documentclass{kluwer}
\usepackage[dvips]{graphicx}
\begin{document}
\begin{article}
\begin{opening}
\title{Palomar Testbed Interferometer Observations of Young Stellar Objects}
        

\author{F.~P.~\surname{Wilkin}\email{f.wilkin@astrosmo.unam.mx}
\thanks{Partially supported by NSF International Researchers Fellows Program}}
\institute{Instituto de Astronomia, 
Universidad Nacional Aut\'onoma de M\'exico, 
Unidad Morelia, Apdo.Postal 3-72 (Xangari), 
58090 Morelia, Michoac\'an, M\'exico}
\author{R.~L.~\surname{Akeson}}
\institute{Michelson Science Center, 
California Institute of Technology, MS 100-22, Pasadena, CA 91125,  USA}

\runningtitle{PTI Observations of YSOs}
\runningauthor{Wilkin \& Akeson}


\begin{abstract} 
We present observations of a sample of Herbig AeBe stars, 
as well as the FU Orionis object V1057 Cygni. Our K-band ($2.2\,\mu m$) 
observations from the Palomar Testbed Interferometer (PTI) used baselines of 
110m and 85m, resulting in fringe spacings of $\sim 4 \,mas$ and $5\, mas$, 
respectively.  
Fringes were obtained for the first time on V1057 Cygni
as well as V594 Cas. Additional measurements were made of MWC147, 
while upper limits to visibility-squared are obtained for  
MWC297, HD190073, and MWC614. These measurements are sensitive to the
distribution of warm, circumstellar dust in these sources. 
If the circumstellar infrared emission comes from warm dust in a disk,
the inclination of the disk to the line of sight implies that the observed
interferometric visibilities should depend upon hour angle. 
Surprisingly, the observations of Millan-Gabet, Schloerb, \& Traub 2001 
(hereafter MST) did not show significant variation  with hour angle. 
However, limited sampling of angular frequencies on the sky 
was possible with the IOTA interferometer, motivating us to study
a subset of their objects  to further constrain these systems. 
\end{abstract}




\end{opening}

\section{INTRODUCTION}
Near-infrared, long baseline interferometry is sensitive to the distribution
of dust around the nearest young stars on scales of the order of 1 AU, and
provides a powerful probe of models of disks and envelopes of such stars. 
The Herbig Ae-Be stars are pre-main sequence, emission line objects that 
are the intermediate mass ($1.5-10\, M_\odot$) counterparts of T Tauri 
stars (Hillenbrand {\it et al.}~1992).
We also observed the FU Orionis object V1057 Cyg, expected to have a
strong disk signature due to the high accretion rate of such objects. 
While the evolutionary status of the FU Orionis objects remains unclear,
they are believed to be T Tauri stars undergoing an episode of greatly
increased disk accretion, involving a brightening of $\sim 5$ magnitudes. 
V1057 Cyg, whose outburst began in 1969-70, is the only FU Orionis
object for which a pre-outburst spectrum is available, confirming its 
pre-main sequence nature (Grasdalen 1973). 
Until now, only one FU Orionis object, FU Orionis itself, has been resolved
by long baseline optical interferometry (Malbet {\it et al.}~1998), and
V1057 Cyg was chosen for study as the next-brightest such object accessible
to PTI. 

\section{SOURCE SELECTION AND OBSERVATIONS}
We selected a sample of 5 sources from the thesis of Millan-Gabet,
chosen to satisfy the observing limitations of PTI, and to avoid
known binaries (with the exception of MWC 147, whose companion is
too faint to affect the current measurements). 
Details of the instrument are described in Colavita 
{\it et al.}~1999.  Table~I describes our final sample. 
%
\begin{table}[h]
\caption{Observing Sample}\label{tab1}
\vskip 0.8cm
\begin{tabular}{llllllll}
\hline \\[-1ex]
{\bf Name} & {\bf Alternate}& {\bf RA ($2000.0$)}& {\bf Dec ($2000.0$)}& {\bf $m_V$ }& {\bf $m_K$}& {\bf Spec}& {d, pc}\\
&&&&&&    \\
\hline \\
HBC 330&  	V594 Cas&  00 43 $18.260$& +61 54 40.100&   9.9&  5.7& B8e&650\\
HD 259431& 	MWC 147&   06 33 $05.190$& $+10$ 19 19.984&   8.7&  5.7& B6pe&800\\
MWC 297&  	NZ Ser&	   18 27 $39.6$  & $-03$ 49 52&       9.5&  3.1& O9e&450\\
HD 179218& 	MWC 614&   19 11 $11.254$& +15 47 15.630&   7.4&  5.9& B9&240\\
HD 190073& 	V1295 Aql& 20 03 $02.510$& +05 44 16.676&   7.8&  5.8& A2pe&280\\
HBC 300&  	V1057 Cyg& 20 58 $53.73$  &   +44 15 28.4&  11.6& 5.7& ---&575\\
\hline
\end{tabular}
\end{table}

Observations of each source were interweaved with nearby calibrator 
stars,  chosen to exclude known binaries and variable stars. 
System visibility was determined based upon observations of the calibrators and 
models of the calibrator (e.g. size based upon multiwavelength photometry). 
The measured raw source visibilities were then
divided by the system visibility. The resulting calibrated visibilities 
$V_{obs}^2$ are presented in Table~II. 
Our reported visibilities are a wideband
average produced synthetically from five narrowband channels. 
As a consistency check,  sources were calibrated first relative 
to one calibrator, then relative to another, and the results compared 
to avoid problems with unknown binarity. The stellar contribution to 
$V_{obs}^2$ is subsequently removed, 
assuming the observed spatial distribution of emission on the sky 
is the sum of an unresolved point source of
known flux, and an extended circumstellar contribution.  
For the Herbig stars, MST estimated the fractions 
of the infrared  emission due to the star and due 
to circumstellar emission at K. 
In Table~II we list the 
fraction $f_{cs,K}$ of emission due to circumstellar matter, while  that of 
the star is $f_{*,K}=1-f_{cs,K}$. 
For V1057 Cyg,  we will assume all the infrared emission  
is circumstellar. 
Table~II also gives $V^2_{cs,K}$ for the  circumstellar contribution, 
where $V_{obs}=f_{*,K}+f_{cs,K} V_{cs}$. 
Because our program stars all have large infrared excesses, 
the corrections for stellar  light are generally small. 
Upper  limits to the visibility squared were determined for sources
lacking fringes, based upon the sensitivity of the detection algorithm 
and measuring the system visibility with a nearby calibrator.
Figures 1-2 show some of the measured individual  
visibilities $V_{obs}^2$ for our resolved sources.
\begin{table}[h]
\caption{Calibrated Visibilities and Stellar Contribution}\label{tab2}
\vskip 0.8cm
\begin{tabular}{lcccccc}
\hline \\[-1ex]
{\bf Source} & {\bf Baseline}&{\bf $V_{obs}^2$} &$f_{cs,K}$&$V_{cs}^2$ \\
 & &  & & \\
\hline \\
V594 Cas	&NW    & $0.40 \pm 0.04$ &$0.92\pm0.03$&$0.36\pm0.04$ \\
MWC 147		&NW    & $0.58 \pm 0.03$ &$0.84\pm 0.04$&$0.51\pm0.04$\\
MWC 147		&NS    & $0.54 \pm 0.04$ &$0.84\pm 0.04$&$0.47\pm0.05$\\
V1057 Cyg	&NW    & $0.63 \pm 0.03$ &$1.000$         &$0.63\pm0.03$\\
MWC 297		&NW,NS & $<0.2 $         &$0.84\pm 0.03$&$<0.12$\\
MWC 614		&NW,NS    & $<0.2 $         &$0.71\pm 0.10$&$<0.05$\\
V1295 Aql	&NS    & $<0.2 $         &$0.72\pm 0.03$&$<0.05$\\
\hline
\end{tabular}
\end{table}
%
\begin{figure}[h]
\begin{center}
\centerline{\resizebox{80mm}{!}{\includegraphics{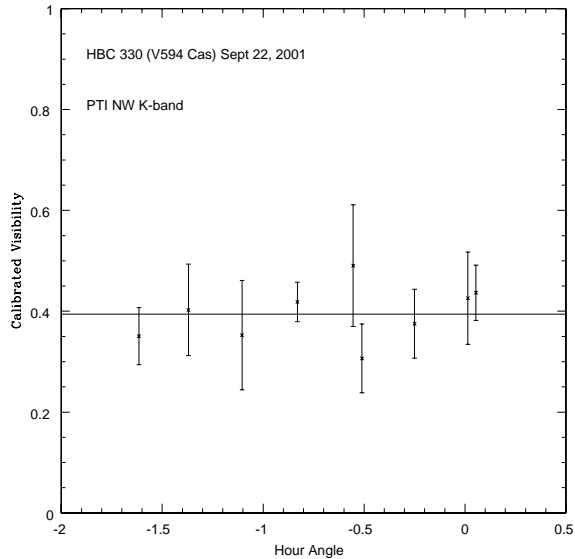}}}
\caption{Calibrated Visibility of HBC 330 (V594 Cas) as a function of 
hour angle. The limited range of hour angles does not permit
strong constaints on the source geometry.}
\label{fig:hbc330}
\end{center}
\end{figure}
%
%
\begin{figure}[h]
\begin{center}
\centerline{\resizebox{80mm}{!}{\includegraphics[width=30pc]{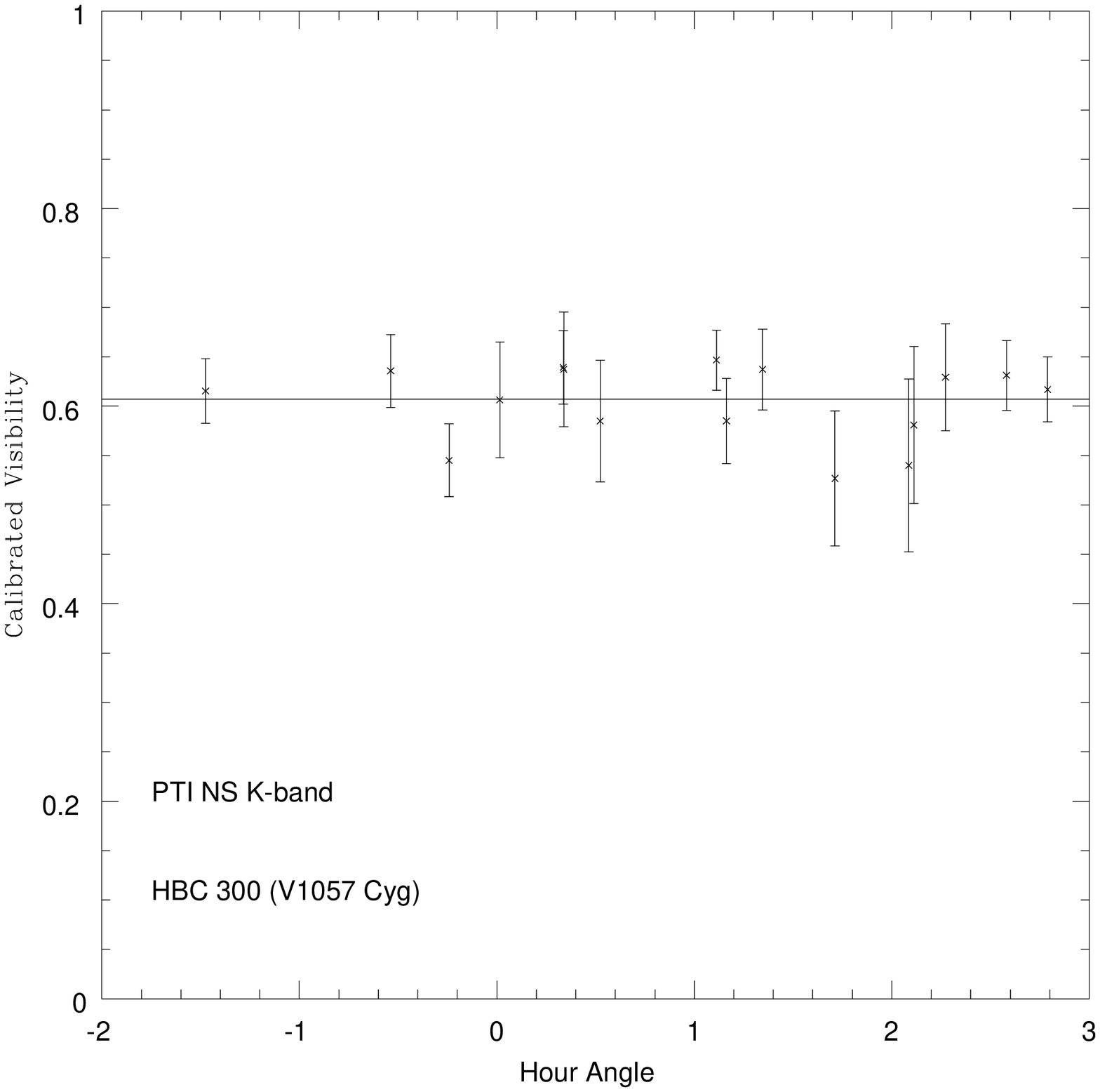}}}
\caption{Calibrated visibility versus hour angle for the FU Orionis 
star V1057 Cyg. The lack of a significant trend suggests a circularly 
symmetric geometry on the sky.}
\label{fig:hbc300}
\end{center}
\end{figure}

Fringes were obtained for a total of four sources, although for one
of these, MWC 297, there are insufficient data to produce a calibrated 
measurement. Thus, we treat MWC 297 as an upper limit. 
Based upon the observed circumstellar visibilities $V_{cs}^2$, 
Table~III gives  approximate source sizes based 
upon a circular Gaussian and a uniform disk model:
\begin{eqnarray} V_{Gauss}^2& =& \Biggl(\exp\Biggl\{-{{\pi^2}\over {\ln\,2}}\,
\biggl({{\theta_{FWHM}}\over {2}}\biggr)^2\,
{{B_p^2}\over {\lambda^2}}\Biggr\}\Biggr)^2,\nonumber\\
 V_{UD}^2& = &\Biggl({{2 J_1(\pi\,\theta_{UD}\,B_p/\lambda)} \over {\pi\,\theta_{UD}\,B_p/\lambda}}
\Biggr)^2.\nonumber
\end{eqnarray}
Here $\lambda=2.235\,\mu m$, 
$B_p$ the projected baseline, $\theta_{FWHM}$ is
the FWHM in radians, $\theta_{UD}$
is the uniform disk diameter in radians, and 
$J_1$ is a Bessel function. 
The baseline lengths  are 110 m in NS, and 85 m in NW. 
Error bars include uncertainties in our measurements and in the stellar and 
circumstellar fluxes, 
but not in the distance. 
\begin{table}[t]
\caption{Source Diameters for the Gaussian and Uniform Disk Models}
\label{tab3}
\vskip 0.8cm
\begin{tabular}{lccccc}
\hline \\[-1ex]
{\bf Source} & {\bf Baseline}&\multicolumn{2}{c}{\bf Gaussian}  
  &\multicolumn{2}{c}{\bf Uniform Disk}   \\
 & &{\bf (mas)}  &{\bf (AU)} &{\bf (mas)} & {\bf (AU)} \\
\hline \\
V594 Cas	&NW    &$2.07\pm 0.11$&$1.35\pm0.07$&$3.34\pm0.16$&$2.17\pm 0.11$\\
MWC 147		&NW    &$1.63\pm 0.10$&$1.30\pm0.06$&$2.75\pm0.15$&$2.20\pm0.12$\\
MWC 147		&NS    &$1.34\pm 0.09$&$1.07\pm0.07$&$2.24\pm0.15$&$1.80\pm0.12$\\
V1057 Cyg	&NW    &$1.36\pm 0.07$&$0.78\pm0.04$&$2.30\pm0.11$&$1.32\pm0.07$\\
MWC 297		&NW    &$>2.9$&$>1.3$&$>4.6 $&$>2.1$\\
MWC 614		&NW    &$>3.5$&$>0.84$&$>5.2$&$>1.2  $\\
V1295 Aql	&NS    &$>2.7$&$>0.76$&$>4.0$&$>1.1$\\
\hline
\end{tabular}
\end{table}

\section{DISCUSSION}
For our observations with the largest
range of hour angles and projected baseline orientation, V1057 Cygni
is consistent with a circularly symmetric source. As an FU Ori type
object, there is little doubt that its  infrared excess comes from
a circumstellar disk and not a spherical distribution of dust. More modeling
is necessary to put limits on the possible orientation of the
disk. 
Our measurements of MWC 147 in the NS baseline are consistent
with those of Akeson et al.~2000. However, the new measurement in the NW 
baseline is inconsisitent with that of the NS baseline if the source
is indeed a circularly-symmetric distribution on the sky. Because the
baselines have differring orientations, the difference can be accounted
for by an asymmetric source distribution, such as a tilted disk. We wish
to confirm the new NW measurement and perform further modeling of this source.

\section{CONCLUSIONS}
We have resolved three young stellar objects at milli-arc second scales,
two for the first time (the Herbig Be star V594 Cas,  and the FU Orionis star 
V1057 Cyg). 
Presumably we are  sampling the distribution of warm dust close to the stars. 
However, these data alone are insufficient to fully constrain the sources,
and other explanations besides circumstellar disks 
(e.g. a binary companion) are possible. 
No significant variation 
of the visibility is seen as a function of hour angle on the sky, 
suggesting a symmetric distribution on the sky. However, for MWC 147,
the measurements in two different baselines suggest an asymmetric 
distribution, such as a tilted disk. This is consistent with recent 
measurements by Eisner {\it et al.}~(2003) for a similar sample 
of Herbig stars, three of which appear to have disks with 
significant inclinations. 

This work has made use of software produced by the Michelson 
Science Center at the California Institute of Technology. F.P.W. is
grateful to the Observatoire de la C\^ote d'Azur for a Poincar\'e
fellowship, and to the NSF International Researchers Fellows Program
for financial support. 
\medskip

\centerline {\bf References}
\vskip 0.4truecm
\noindent
Akeson, R.L., Ciardi, D.R., van Belle, G.T., Creech-Eakman, M.J., \\
\& Lada, E.A. 2000, ApJ, 543, 313\\
Colavita, M.M., Wallace, J.K., Hines, B.E., {\it et al.}~1999, ApJ, 510, 505\\
Eisner, J.A., Lane, B.F., Akeson, R.L., Hillenbrand, L.A., \& Sargent, A.I. 2003, ApJ, (in press)\\
Grasdalen, G.L. 1973, ApJ, 182, 781\\
Hillenbrand, L.A., Strom, S.E., Vrba, F.J., \& Keene, J. 1992, ApJ, 397, 613\\
Malbet, F., Berger, J.-P., Colavita, M.M., {\it et al.}~2000, ApJ, 543, 313\\
Millan-Gabet, R., Schloerb, F.P., \& Traub, W.A. 2001, ApJ, 546, 358\\

\end{article}
\end{document}